%% file: main.tex
\documentclass{article}

\usepackage[english]{babel}

\usepackage[letterpaper,top=2cm,bottom=2cm,left=3cm,right=3cm,marginparwidth=1.75cm]{geometry}

\usepackage{amsmath}
\usepackage{graphicx}
\usepackage[colorlinks=true, allcolors=blue]{hyperref}

\usepackage{booktabs} 
\usepackage{array}
\usepackage{subcaption}
\usepackage[sorting=none,style=nature]{biblatex}
\usepackage{dirtree}
\usepackage[most]{tcolorbox}
\usepackage{authblk}
\title{ULMShare: A Large-Scale In Vivo Ultrasound Localization Microscopy Dataset for Microvascular Imaging}
\author[1,2]{Brice Rauby\textsuperscript{\textdagger}}
\author[1]{Nin Ghigo\textsuperscript{\textdagger}}
\author[3]{Gerardo Ramos-Palacios}
\author[1]{Alexis Leconte}
\author[1]{Stephen A. Lee}
\author[1]{Alice Wu}
\author[1]{Paul Xing}
\author[1]{Oleksandra Gulenko}
\author[1]{Louis Caron}
\author[4,5]{Antoine Malescot}
\author[4,5]{Eric Martineau}
\author[1]{Jonathan Porée}
\author[2,6,7]{Maxime Gasse}
\author[5,8,9]{Ravi L. Rungta}
\author[3]{Abbas F. Sadikot}
\author[1,9]{Jean Provost\textsuperscript{*}}

\affil[1]{\small Department of Engineering Physics, Polytechnique Montréal, Montreal, QC, Canada}
\affil[2]{\small Mila - Quebec Artificial Intelligence Institute, Montreal, QC, Canada}
\affil[3]{\small Montreal Neurological Institute, McGill University, Montreal, QC, Canada}
\affil[4]{\small Department of Physiology and Pharmacology, Université de Montréal, QC, Canada}
\affil[5]{\small Department of Stomatology, Université de Montréal, QC, Canada}
\affil[6]{\small Microsoft AI, Montreal, QC, Canada}
\affil[7]{\small Department of Computer Engineering and Software Engineering, Polytechnique Montréal, Montreal, QC, Canada}
\affil[8]{\small Department of Neuroscience, Université de Montréal, QC, Canada}
\affil[9]{\small Centre Interdisciplinaire de Recherche sur le Cerveau et l'Apprentissage, Université de Montréal, QC, Canada}
\affil[10]{\small Montreal Heart Institute, Montreal, QC, Canada}

\date{}

\begin{document}
\maketitle
\begin{center}
    \footnotesize
    $^{*}$ Corresponding author: jean.provost@polymtl.ca \\
    $^{\dagger}$ These authors contributed equally to this work.
\end{center}
\input{sections/abstract}
\input{sections/Background_Summary}
\input{sections/methods}
\input{sections/Data_Record}
\input{sections/Technical_Validation}
\input{sections/Usage_Notes}
\section{Data Availability}
The raw dataset, comprising approximately 30 TB of radio-frequency data and metadata from 99 acquisitions, is available via the \href{https://doi.org/10.20383/103.01550}{Federated Research Data Repository (FRDR)}. Derived ULM maps, microbubble trajectories, and MATLAB helper scripts are hosted on the \href{https://github.com/provostultrasoundlab/ulmshare}{companion GitHub repository}.

\input{sections/Code_Availability}
\section{Funding}
This work was supported in part by the Institute for Data Valorization (IVADO), in part by the Canada Foundation for Innovation under Grant 38095, in part by the Canadian Institutes of Health Research (CIHR) under Grant 452530, in part by the New Frontiers in Research Fund under Grant NFRFE-2018-01312, in part by the Natural Sciences and Engineering Research Council of Canada (NSERC) under Grant (RGPIN-2019-04982 and RGPIN-2020-05276), and in part Canada Research Chair in neurovascular interactions. Further support came from the TransMedTech Institute, the Fonds de recherche du Québec - Nature et technologies, the Quebec Bio-Imaging Network, the CONAHCYT, the Insightec, the Healthy Brains Healthy Lives, the Canada First Research Excellence Fund, and the NSERC. Additionally, computational resources and data storage were provided through the Digital Research Alliance of Canada. 
\printbibliography 

\end{document}

%% file: sections/abstract.tex
\begin{abstract}
Ultrasound Localization Microscopy (ULM) enables microscopic imaging of the cerebral microvasculature \textit{in vivo}, but relies on a multi-stage processing pipeline in which acquisition settings and reconstruction processes strongly influence the final output. 
Existing public datasets remain sparse, restricting rigorous evaluation and slowing progress in algorithm development, including emerging machine-learning approaches, which by design require large quantities of data to be robust and reliable.
We introduce \textbf{ULMShare}, an open-access dataset of 99 whole-brain transcranial ULM acquisitions from 61 healthy mice (36 females, 22 males, 3 unknown; mean age: $8.2 \pm 5.5$ weeks; mean weight: $17.7 \pm 4.2$ g), for a total of 30TB of raw data. The dataset spans three experimental procedures, multiple injection and anesthesia protocols, two ultrasound probes, and different imaging planes and orientations. Each acquisition includes raw ultrasonic data, detailed metadata, an illustrative reconstruction and the corresponding microbubble trajectories.
Alongside the data, we report vascular saturation, Fourier Ring Correlation, and track-length statistics, plus expert visual gradings. ULMShare provides a broad, standardized and publicly available resource for method development, validation, and benchmarking. The full dataset is  available on the \href{https://doi.org/10.20383/103.01550}{Federated Research Data Repository} and additional resources are hosted on the \href{https://github.com/provostultrasoundlab/ulmshare}{ULMShare Github repository}.
\end{abstract}

%% file: sections/Background_Summary.tex
\section{Background \& Summary}
Ultrasound Localization Microscopy (ULM) is a super‑resolution ultrasound technique that images microvasculature \textit{in vivo}, at depth, and at resolutions breaching the diffraction limit \cite{errico_ultrafast_2015,couture_ultrasound_2018, christensen-jeffries_super-resolution_2020}. By localizing and tracking clinically approved microbubbles (MB) injected into the bloodstream, ULM reconstructs detailed maps of the microvascular network and provides quantitative measurements of blood flow dynamics.

In the brain, ULM studies on animal models have been critical to develop new sequences and methodologies~\cite{bourquin_vivo_2021} enabling extraction of complex hemodynamic markers such as pulsatility imaging~\cite{bourquin_vivo_2021, bourquin_quantitative_2024, ghigo_dynamic_2024}, or functional imaging \cite{renaudin_functional_2022,shin_context-aware_2024}, as well as for resolving capillary-level vasculature \cite{lee_functional_2024,heiles_nonlinear_2024}.

Despite these promising advances, ULM remains a modality that relies on experimental protocols requiring costly equipment and skilled technicians, which may induce high variability in acquisition quality even within the same protocol. Consequently, new  methodological developments are often evaluated on a reduced number of \textit{in vivo} acquisitions, which limits their generalization and overall impact. Additionally, some groups lack access to the required equipment and are therefore unable to validate their methods on \textit{in vivo} data.

ULM relies on a complex reconstruction pipeline: beamforming, clutter filtering, MB localization and tracking—where each step can significantly impact the output quality. Open-access datasets are critical as they facilitate fair comparison between methodologies and improve reproducibility. Recent efforts were made within the community to release open-access datasets such as PALA \cite{heiles_performance_2022} or companion datasets to animal studies \cite{renaudin_functional_2022,denis_sensing_2023,shin_context-aware_2024,bureau_ultrasound_2025,chabouh_3d_2025}.
The PALA benchmark has gained substantial traction~\cite{sui_generative_2022,wang_ultrasound_2022,corazza_microbubble_2022,zhang_transformer_2022,pham_performance_2023,hahne_geometric_2023,corazza_microbubble_2023,liu_ultrasound_2023,corazza_adaptive_2023,hahne_learning_2023,hahne_stofnet_2023,xiao_deep-learning-based_2023,tuccio_relaxing_2023,pialot_simplified_2023,afrakhteh_novel_2024,  lan_deep_2024,tuccio_time_2024,hahne_rf-ulm_2024,zhang_ulm-mbcnrt_2024, chen_competitive_2024,tuccio_towards_2024,qiang_adaptive_2024,fang_total_2024,li_attention_2024,,zhang_efficient_2024,han_tissue_2025,pustovalov_computational_2025,tong_dual_2025,pham_leveraging_2025,pialot_computationally_2025,cho_numerical_2025} demonstrating the strong demand for publicly available data to support the development and validation of ULM processing algorithms. However, existing public datasets remain scarce and are mostly limited to simulated data or a small number of \textit{in vivo} acquisitions~\cite{heiles_performance_2022}.

Large and diverse \textit{in vivo} ULM datasets are essential for statistically rigorous method evaluation under varied protocols, reproducibility, fair benchmarking, and the development of robust machine-learning models~\cite{raubyDeepLearningUltrasound2024a}. By lowering the barrier to entry and enabling researchers without access to costly equipment to work directly with \textit{in vivo} data, such resources can accelerate innovation and broaden the impact of ULM across the scientific community. 

In this paper, we present \textbf{ULMShare}, an open-access dataset of  ultrasound acquisitions, unique by its scale, two orders of magnitude larger than existing public ULM dataset. 99 acquisitions, each lasting several minutes, were acquired in 61 anesthetized wild-type mice (aged 3 to 16 weeks) through intact skin and skull, totaling over 30 TB. The full dataset is made available publicly as collection on the \href{https://doi.org/10.20383/103.01550}{Federated Research Data Repository} and ULM reconstructions and helper scripts are provided on a companion \href{https://github.com/provostultrasoundlab/ulmshare}{Github repository}.

ULMShare comprises both previously unpublished data and data that have been used in prior publications.
Continuous transcranial ULM acquisitions were first presented in Lee et al. \cite{lee_functional_2024}, where they were used to demonstrate temporally continuous imaging for functional brain mapping.  
Data from Xing et al. \cite{xing_phase_2024, xing_inverse_2025} were employed to investigate aberration correction techniques, highlighting methods to improve image quality and localization accuracy in transcranial ULM.  
Acquisitions reported in Ghigo et al. \cite{ghigo_dynamic_2024} were used for pulsatility evaluation, demonstrating the potential of ULM to capture dynamic vascular signals.  
Finally, datasets included in Leconte et al. \cite{leconte_tracking_2024} supported the development and validation of a spatiotemporal tracking algorithm for MB trajectories.  

%% file: sections/methods.tex
\section{Methods}
\subsection{ULM acquisition}

\begin{figure}[htbp]
    \centering
    \includegraphics[width=0.9\linewidth]{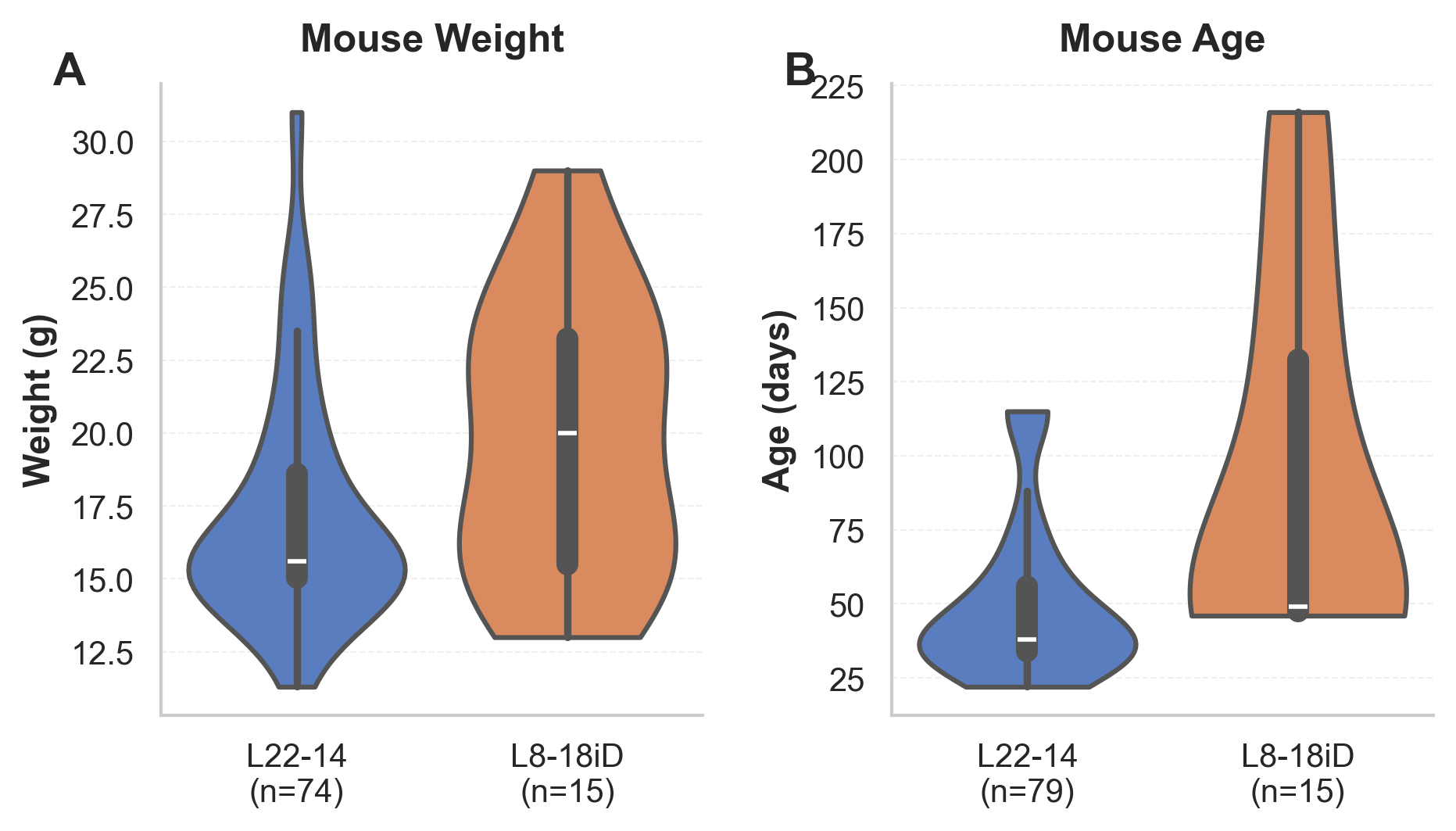}
    \caption{Overview of animal-specific information.}
    \label{fig:animalOverview}
\end{figure}

ULMShare consists of \textbf{99} transcranial brain ULM acquisitions obtained from 61 healthy mice (36 females, 22 males, 3 unknown; mean age: $8.2 \pm 5.5$ weeks; mean weight: $17.7 \pm 4.2$ g; see Fig.~\ref{fig:animalOverview}). Experimental procedures were approved by the Animal Care Ethics Committees of the University of Montreal (22-013), McGill University (AUP-4532), and the Montreal Heart Institute (2023-32-02 TAC-ultrasons).

\subsubsection{Data Acquisition Protocols}
The dataset aggregates acquisitions collected over three years using three distinct experimental protocols, previously described in published articles \cite{ghigo_dynamic_2024,leconte_tracking_2024,lee_functional_2024,xing_inverse_2025,xing_phase_2024}. 

Common to all procedures, animals were secured in a stereotaxic frame to minimize motion. For the majority of acquisitions ($n=97$), the scalp was shaved and degassed ultrasound gel was applied.  For 2 acquisitions (Protocol 3), the scalp was surgically removed to minimize attenuation. All data were acquired using a Vantage-256 system (Verasonics, WA, USA) emitting tilted plane waves ($-5^\circ$ to $5^\circ$, centered at $0^\circ$). The specific acquisition parameters differentiating the three protocols are detailed in Table~\ref{tab:protocols}.

\begin{table}[ht]
    \centering
    \caption{Comparison of Acquisition Protocols used in ULMShare.}
    \label{tab:protocols}
    \begin{tabular}{lccc}
        \toprule
        \textbf{Parameter} & \textbf{Protocol 1} & \textbf{Protocol 2} & \textbf{Protocol 3} \\
        \midrule
        Reference & \cite{ghigo_dynamic_2024, leconte_tracking_2024} & \cite{lee_functional_2024} & \cite{xing_inverse_2025,xing_phase_2024} \\
        \textbf{Acquisitions ($n$)} & \textbf{71} & \textbf{15} & \textbf{13} \\
        Anesthesia & Isoflurane (1--2\%) & Isoflurane (1--2\%) & Ketamine/Medetomidine \\
        MB Injection & Tail vein injection & Tail vein catheter \textsuperscript{a} & Retro-orbital injection \\
        Transducer & L22-14v & L8-18iD & L22-14v \\
        Center Freq. & 15.625 MHz & 10.4667 MHz & 15.625 MHz \\
        Plane Waves & 11 & 9 & 11 \\
        Frame Rate & 1000 Hz & 1000 Hz & 1000 or 1600 Hz \\
        Duration & 160 s & 300 s & 300 s \\
        Sampling & 100\% Bandwidth & 50\% Bandwidth & 100\% Bandwidth \\
        \bottomrule
        \multicolumn{4}{l}{\footnotesize \textsuperscript{a} \textit{Note: 3 acquisitions followed Protocol 2 but used tail vein injection.}} \\
    \end{tabular}
\end{table}

\subsubsection{Dataset Composition and Variations}
The 99 total acquisitions include specific subsets designed for sensitivity analysis and anatomical mapping, which are included within the protocol counts listed in Table~\ref{tab:DetailedComposition}. All the following details are systematically provided in the metadata.

Regarding Protocol 1, while the previously published configuration used a 1:10 saline dilution ($n=6$), the majority of acquisitions employed a 1:50 dilution ($n=64$); one acquisition used a 1:20 dilution. Additionally, sequential acquisitions were performed in 3 mice to image different anatomical planes, yielding 17 acquisitions (5--6 per mouse) taken at 0.5 mm intervals.
Protocol 2 includes 3 acquisitions that combined the animal preparation of Protocol 1 (tail vein injection, 1:20 dilution, 4~$\mu$L/g) with the imaging sequence of Protocol 2 (L8-18iD probe, 300 s duration).
In Protocol 3, a subset of acquisitions was collected during stimulation for functional imaging (treated here as resting-state due to lack of response), and for 2 acquisitions (1 mouse), the scalp was surgically removed to minimize attenuation.

Finally, for 23 acquisitions across the dataset, an expert manually identified the anatomical slice position, including striatum ($n=11$), pons ($n=2$), midbrain ($n=4$), hippocampus ($n=3$), and cerebellum ($n=3$). Detailed metadata for every acquisition is provided in the accompanying JSON files.

\begin{table}[htbp]
\centering
\caption{Detailed composition of the ULMShare dataset grouped by experimental protocol.}
\label{tab:DetailedComposition}
\resizebox{\textwidth}{!}{%
\begin{tabular}{c c l c c c l}
\toprule
\textbf{Mice} & \textbf{Acq.} & \textbf{Injection Route} & \textbf{Dilution} & \textbf{Volume} & \textbf{Probe} & \textbf{Anatomical Coverage (Acquisitions)} \\
\midrule
\multicolumn{7}{l}{\textit{\textbf{Protocol 1} \cite{ghigo_dynamic_2024, leconte_tracking_2024}}} \\
36 & 64 & Tail vein & 1:50 & 4 $\mu$L/g & L22-14v & Striatum ($n=11$), Midbrain ($n=4$), Hippo. ($n=3$), \\
   &    &           &      &            &         & Cerebellum ($n=3$), Pons ($n=2$), Unreg. ($n=41$) \\
3 & 6 & Tail vein & 1:10 & 4 $\mu$L/g & L22-14v & Unregistered \\
1 & 1 & Tail vein & 1:20 & 4 $\mu$L/g & L22-14v & Unregistered \\

\midrule
\multicolumn{7}{l}{\textit{\textbf{Protocol 2} \cite{lee_functional_2024}}} \\
12 & 12 & Tail vein catheter & 1:10 & 4 $\mu$L/g & L8-18iD & Unregistered \\
3 & 3 & Tail vein & 1:20 & 4 $\mu$L/g & L8-18iD & Bregma $-3$ mm ($n=3$) \\

\midrule
\multicolumn{7}{l}{\textit{\textbf{Protocol 3} \cite{xing_inverse_2025}}} \\
5 & 11 & Retro-orbital & Pure & 10 $\mu$L & L22-14v & Functional / Deep Brain targets \\
1 & 2 & Retro-orbital & Pure & 10 $\mu$L & L22-14v & Functional (Scalp Removed) \\

\bottomrule
\multicolumn{7}{l}{\footnotesize \textbf{Total:} 61 Mice, 99 Acquisitions.} \\
\end{tabular}%
}
\end{table}
\subsection{ULM Processing}
In addition to the raw data, ULMShare provides for each acquisition an illustrative signed ULM density map and the corresponding detected microbubble (MB) tracks. These ULM maps were generated using fixed, non-optimized processing parameters. Their purpose is to provide an example of the vascular information that can be obtained from each acquisition; they should not be used as benchmarks for quantitative comparison.
In phase-quadrature data were beamformed onto an isometric grid with a spacing of $\lambda/4$ using standard GPU-based delay-and-sum beamforming. An f-number of 1.4 was used for data acquired with the L22-14v probe, and 1.0 for acquisitions using the L8-18iD probe. Tissue signals were removed from the raw data using Singular Value Decomposition (SVD) clutter filtering \cite{demene_spatiotemporal_2015}. The first 20 singular values were removed for L22-14v acquisitions and the first 30 for L8-18iD acquisitions.

Two signal-enhancement processes were then applied to the clutter-free data \cite{xing_3d_2025}: spatially dependent time-gain compensation (TGC) and lag-1 autocorrelation.  
The spatially dependent TGC aims to improve signal quality in shadowed regions. A 2D Gaussian filter with a standard deviation of $2.25\lambda$ (corresponding to 9 pixels of the beamforming grid) was applied to the Power Doppler (time average of the power across a set of hundred of frames) to compute a spatially dependent TGC map. This map represents an estimate of the spatial intensity distribution and is used to equalize intensity and reduce shadowing artifacts. 
Temporal lag-1 autocorrelation was then applied, reducing noise while enhancing MB signals.

On the resulting MB-enhanced data, a spatiotemporal tracking algorithm was employed to track MB trajectories in space and time \cite{leconte_tracking_2024}. Briefly, MBs moving through space and time resemble tubular structures with radii comparable to the point spread function (PSF), which can be enhanced using a spatiotemporal filter. The tubular radius was set to $\lambda$ across the image. A thinning algorithm was then applied to segment the centerlines of these tubular structures at pixel resolution \cite{wagner_real-time_2020}. Finally, subwavelength MB positions were estimated using a radial symmetry algorithm on regions of interest (ROI) of 15 pixels \cite{heiles_performance_2022}. Tracks shorter than 20 frames were considered noise and discarded.  
Following \cite{leconte_tracking_2024}, $\epsilon$ was fixed at 1.4 to reject tubular structures orthogonal to the temporal dimension, and $\tau$ was fixed at 0.6 to adapt the sensitivity of the function to the non-homogeneous intensity of MB PSF trajectories.  
All ULM processing parameters are summarized in Table \ref{tab:parameters_ULM}.

 \begin{table}[htbp]
\centering
\caption{Summary of fixed parameters used in ULM processing for each probe type.}
\label{tab:parameters_ULM}
\begin{tabular}{|p{4cm}|p{5.2cm}|p{5.2cm}|}
\hline
\textbf{Parameter} & \textbf{L22-14v} & \textbf{L8-18iD} \\
\hline
Beamforming & 
Grid spacing $\lambda/4$ \newline
f-number = 1.4 &
Grid spacing $\lambda/4$ \newline
f-number = 1.0 \\
\hline
Clutter filtering (SVD) &
First 20 singular values removed &
First 30 singular values removed \\
\hline
TGC filter size &
2D Gaussian with $\sigma = 9\lambda$ &
2D Gaussian with $\sigma = 9\lambda$ \\
\hline
Tracking parameters &
Tubular radius = $\lambda$ \newline
$\epsilon = 1.4$, $\tau = 0.6$ &
Tubular radius = $\lambda$ \newline
$\epsilon = 1.4$, $\tau = 0.6$ \\
\hline
ROI for radial symmetry localization &
15 pixels & 
15 pixels \\
\hline
Minimum number of frames per track &
20 & 
20 \\
\hline
\end{tabular}
\end{table}



%% file: sections/Data_Record.tex
\section{Data Records}
The \textsc{ULMShare} dataset is publicly available through the
\href{https://doi.org/10.20383/103.01550}{Federated
 Research Data Repository} (FRDR).
The collection includes 99 transcranial ultrasound acquisitions from 61 mice recorded between March~2022 and March~2025.
Each record corresponds to a single acquisition and contains raw ultrasonic channel data and structured metadata.
In total, the dataset represents approximately 30~TB of raw data.
Some acquisitions correspond to datasets previously used in published studies; their associated DOIs are included in the acquisition-level metadata files and in the global \texttt{metadata.csv} file.

To support transparent processing and reproducibility, we provide example ULM maps, detected microbubble trajectories, and postprocessing metrics for each dataset. These examples are hosted in a dedicated GitHub \href{https://github.com/provostultrasoundlab/ulmshare}{repository} to enable accessibility, version control, and continuous updates, together with a dataset summary and MATLAB helper scripts.

\subsection{Data directory}
\begin{figure}[h]
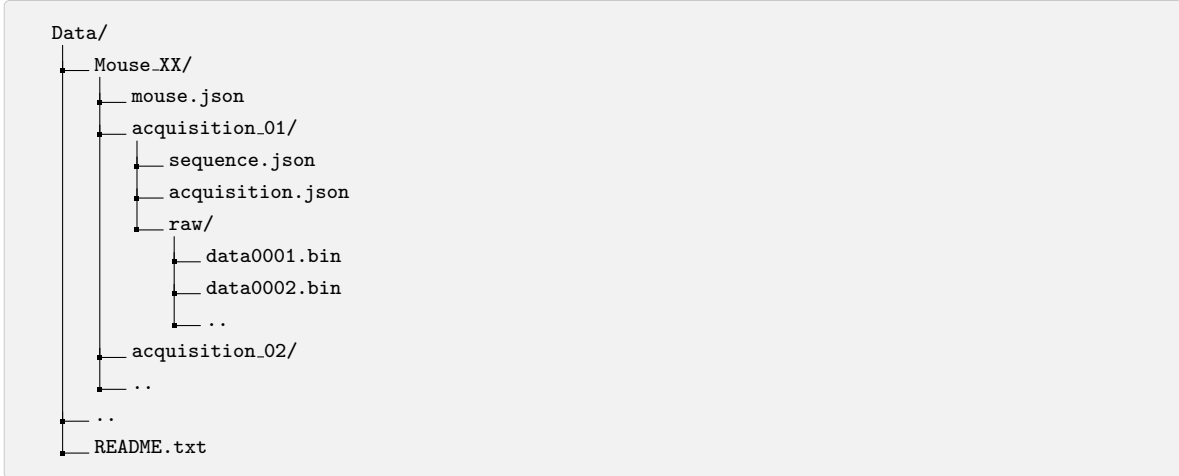

\centering
\begin{tcolorbox}[
    colback=gray!10,
    colframe=gray!50,
    arc=2pt,
    boxrule=0.3pt,
    left=3pt,
    right=3pt,
    top=3pt,
    bottom=3pt
]
\footnotesize
\dirtree{%
.1 Data/.
.2 Mouse\_XX/.
.3 mouse.json.
.3 acquisition\_01/.
.4 sequence.json.
.4 acquisition.json.
.4 raw/.
.5 data0001.bin.
.5 data0002.bin.
.5 ...
.3 acquisition\_02/.
.3 ...
.2 ...
.2 README.txt.
}
\end{tcolorbox}
\caption{Directory structure of the raw data stored in the FRDR repository.}
\label{fig:data_tree_frdr}
\end{figure}

All raw acquisitions are stored under the \texttt{Data/} directory and organized by mouse index, as shown in Fig.~\ref{fig:data_tree_frdr}.
Each mouse folder contains a \texttt{mouse.json} file describing strain, sex, age, body weight, protocol number, and facility location.
Each acquisition directory includes a \texttt{sequence.json} file specifying the imaging parameters (e.g., transducer type, number of plane-wave angles), an \texttt{acquisition.json} file containing acquisition-specific information such as injection parameters or procedure details, and a \texttt{raw/} folder that stores the Verasonics channel data files (\texttt{dataXXXX.bin}).

\subsection{Code directory}

\begin{figure}[h]
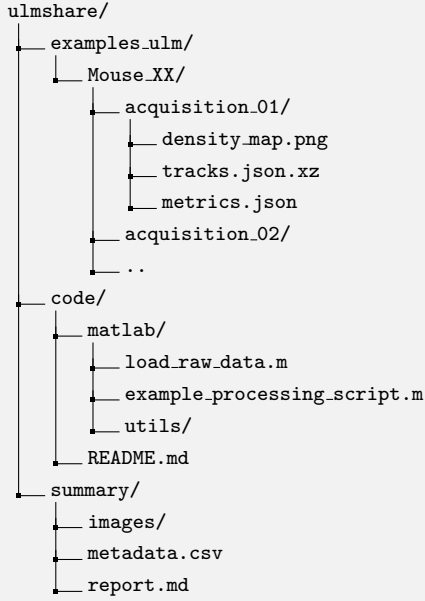

\centering
\begin{tcolorbox}[
    colback=gray!10,
    colframe=gray!50,
    arc=2pt,
    boxrule=0.3pt,
    left=3pt,
    right=3pt,
    top=3pt,
    bottom=3pt
]
\footnotesize
\dirtree{%
.1 ulmshare/.
.2 examples\_ulm/.
.3 Mouse\_XX/.
.4 acquisition\_01/.
.5 density\_map.png.
.5 tracks.json.xz.
.5 metrics.json.
.4 acquisition\_02/.
.4 ...
.2 code/.
.3 matlab/.
.4 load\_raw\_data.m.
.4 example\_processing\_script.m.
.4 utils/.
.3 README.md.
.2 summary/.
.3 images/.
.3 metadata.csv.
.3 report.md.
}
\end{tcolorbox}
\caption{Directory structure of ULM examples, summary, and helper codes hosted in the GitHub repository.}
\label{fig:data_tree_github}
\end{figure}

All example outputs and supporting resources are hosted in a dedicated GitHub
\href{https://github.com/provostultrasoundlab/ulmshare}{repository} to enable accessibility, version control, and future updates.
The repository mirrors the mouse/acquisition hierarchy of the FRDR archive, as shown in Fig.\ref{fig:data_tree_github}.

The \texttt{examples\_ulm/} directory replicates the FRDR mouse/acquisition structure.
Each acquisition folder includes:
\begin{itemize}
    \item an illustrative ULM density map (\texttt{density\_map.png}),
    \item detected microbubble trajectories (\texttt{tracks.json.xz}),
    \item metrics (\texttt{metrics.json}) computed for technical validation as described in \ref{sec:technical_validation}.
\end{itemize}
The file \texttt{tracks.json} contains localized microbubble trajectories without interpolation.
It includes a \texttt{metadata} section (pixel size, frame rate, units, acquisition identifier) and an \texttt{all\_tracks} list, where each element corresponds to a processed buffer with its \texttt{buffer\_index} and associated tracks.

The \texttt{code/} directory provides lightweight MATLAB scripts demonstrating the typical workflow from raw Verasonics buffers to a simple ULM reconstruction as explained in section \ref{sec:code_availability}.
These examples rely on open-source libraries such as the MATLAB UltraSound Toolbox (MUST)~\cite{cigier_simus_2022,garcia_make_2021} and the Tracking and Localization toolbox~\cite{leconte_tracking_2024}.

The \texttt{summary/} directory contains high-level overview material to facilitate navigation:
\begin{itemize}
    \item an aggregated \texttt{metadata.csv} listing all mice and acquisitions,
    \item an illustrative density map in \texttt{images/} for rapid browsing,
    \item a human-readable \texttt{report.md} documenting data organization and provenance.
\end{itemize}

This structure allows users to explore example outputs, inspect metadata without downloading the raw FRDR archive.

%% file: sections/Technical_Validation.tex
\section{Technical Validation}
\label{sec:technical_validation}
To assess the technical quality and reproducibility of ULMShare, we performed  quantitative and qualitative evaluations across all acquisitions. These analyses help readers evaluate the suitability of each acquisition for their intended applications. In addition to many high-quality examples, the dataset intentionally includes acquisitions of varying quality, reflecting the realistic and heterogeneous conditions of transcranial imaging.  This diversity ensures that ULMShare offers both high-quality examples for benchmarking and more challenging cases for the development of robust reconstruction methods. 

\begin{figure}[htbp]
    \centering
    \includegraphics[width=\textwidth]{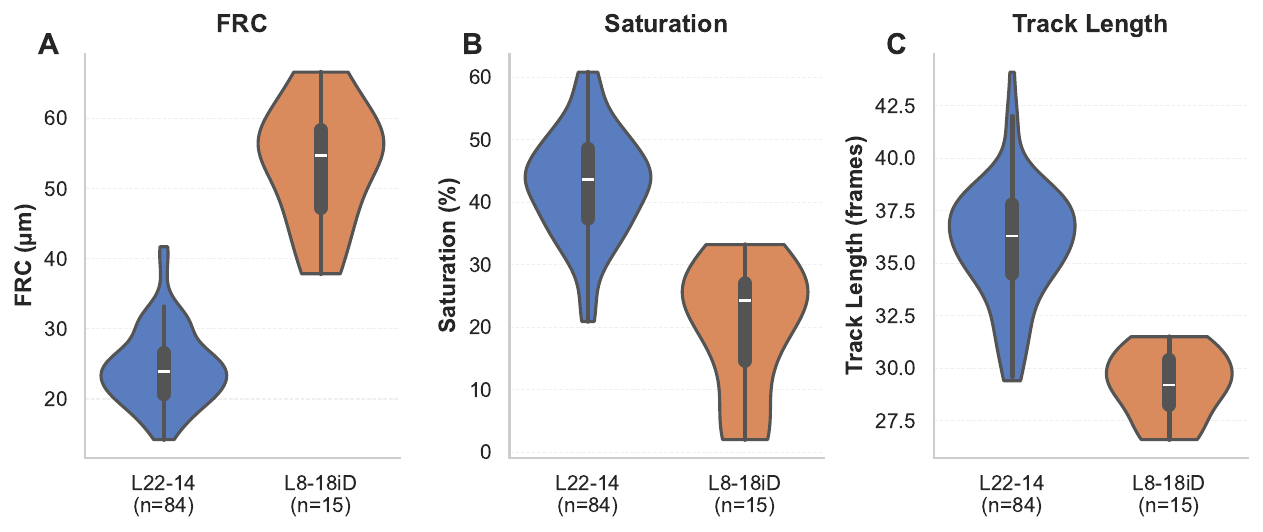}
    \caption{Distribution of ULM validation quantitative metrics across probe types. (A) FRC, (B) vascular saturation, and (C) mean track length. Data shown for L22-14 (n=84) and L8-18iD (n=15) probes.The white dot indicating the median, the thick black bar representing the interquartile range, and the thin black line extending to the data extremes.}
    \label{fig:validation_metrics}
\end{figure}

We evaluated vascular saturation as an indicator of the proportion of perfused vasculature, estimated spatial coherence using Fourier Ring Correlation (FRC) \cite{nieuwenhuizen_measuring_2013}, reported  average length of detected tracks, and carried out visual inspection of all super-resolution maps. Together, these metrics provide a comprehensive assessment of acquisition quality and consistency across the dataset. FRC, vascular and mean track length distributions across the dataset are represented in Figure \ref{fig:validation_metrics}.

\subsection{Vascular saturation}
Vascular saturation quantifies the fraction of the field of view visited by localized MBs throughout an acquisition. It provides a global measure of microvascular sampling density and reflects MB injection efficiency, perfusion, imaging depth, and potential shadowing from the skull.

Across the dataset, saturation values reflected expected differences between probe types. For the L22-14 probe (n=84), vascular saturation ranged from 20.9\% to 61.5\% (median 43.9\%). For the L8-18iD probe (n=15), saturation values were lower, ranging from 0.5\% to 33.2\% (median 24.2\%).

\subsection{FRC - Spatial coherence}

Spatial coherence was evaluated using FRC, which measures the frequency-dependent agreement between two statistically independent reconstructions. For each acquisition, the localized MBs were randomly split into two halves, reconstructed independently, and the FRC curve computed following Hingot \emph{et al.}~\cite{hingot_measuring_2021}. We report the spatial frequency at which the FRC curve crosses the half-bit threshold, which provides a data-driven estimate of the highest spatial frequency that is coherently sampled.

Across the dataset, FRC half-bit spatial coherence scales differed markedly between probes. For the L22-14 probe, values ranged from 14.21 to 41.70 $\mu$m (median 23.71$\mu$m $\sim\lambda/4$). For the L8-18iD probe, coherence scales were broader, ranging from 40.16 to 66.60 $\mu$m (median 54.74 $\mu$m $\sim \lambda /3$). This change reflects the larger wavelength associated with the lower center frequency.

\subsection{Track length - Temporal coherence}
Across acquisitions, track-length statistics exhibited moderate variability. For the L22-14v probe, mean track lengths ranged from 29.40 to 44.10 frames (median 36.30), whereas for the L8-18iD probe they ranged from 26.80 to 31.50 frames (median 29.20). Longer and more continuous trajectories generally indicate higher confidence in MB localization, as they require consistent detection across consecutive frames and are less likely to arise from noise, clutter, or spurious localizations \cite{lee_functional_2024}. 

\subsection{Qualitative inspection}
\begin{figure}[htbp]
\centering

\newcommand{\densheight}{3.5cm}

\begin{subfigure}[b]{0.32\linewidth}
    \centering
    \includegraphics[height=\densheight,trim=40 10 40 10,clip]
        {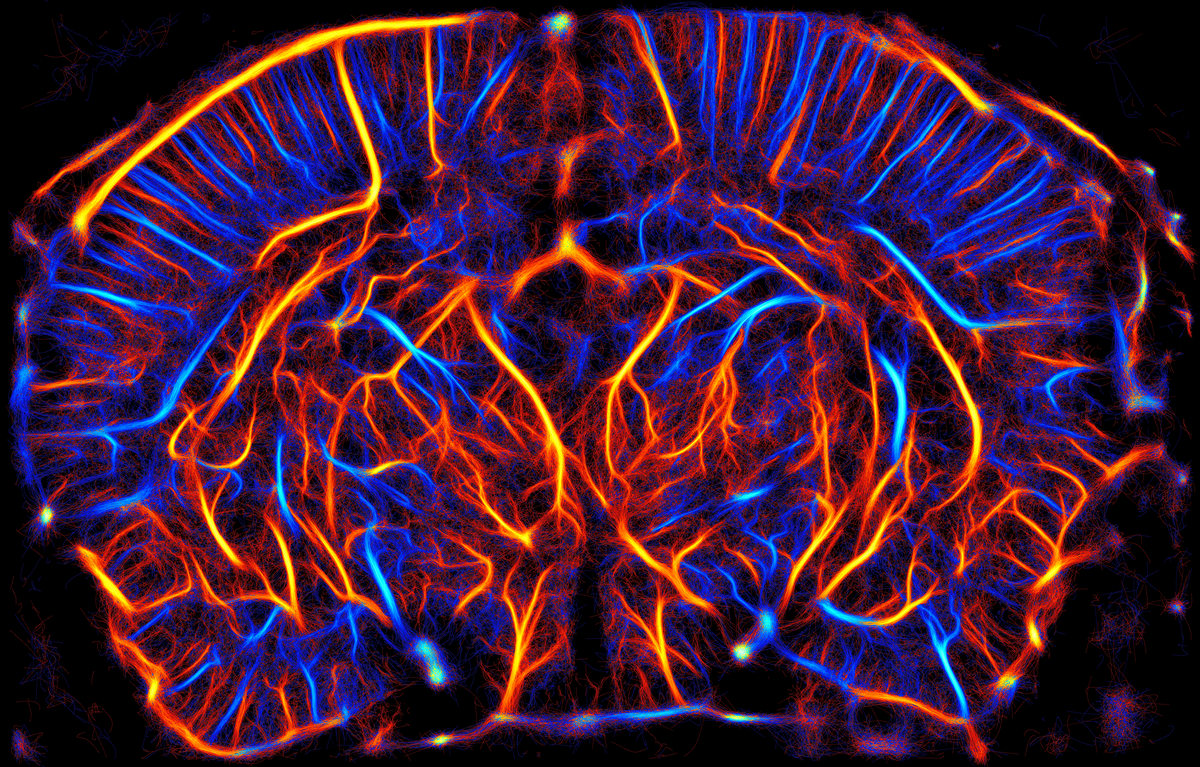}
    \caption*{Grade A}
\end{subfigure}
\begin{subfigure}[b]{0.32\linewidth}
    \centering
        \includegraphics[height=\densheight,trim=90 10 90 10,clip]
        {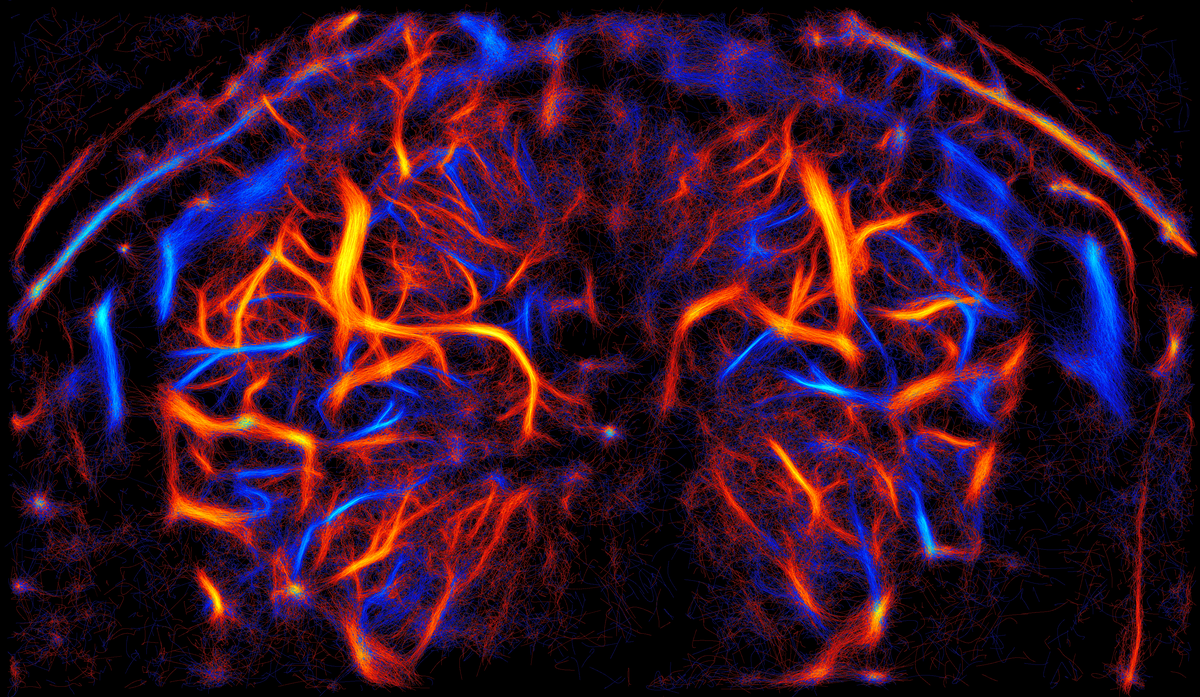}

    \caption*{Grade B}
\end{subfigure}
\begin{subfigure}[b]{0.32\linewidth}
    \centering
    \includegraphics[height=\densheight,trim=90 10 90 10,clip]
        {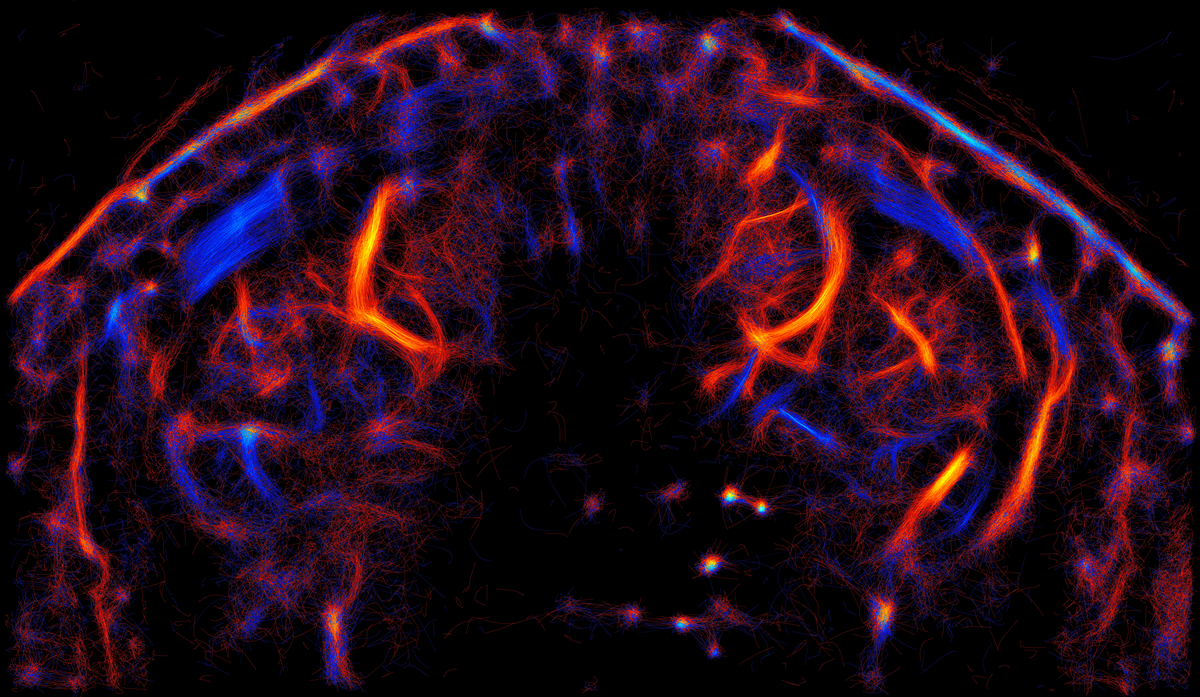}
    \caption*{Grade C}
\end{subfigure}

\caption{
Representative examples of ULM reconstructions (red: flowing downward; blue: flowing upward) illustrating the range of acquisition quality in the ULMShare dataset.
}
\label{fig:visual_grading}
\end{figure}
All acquisitions were visually inspected by two experienced researchers and assigned to a three-level quality scale based on the visibility of fine vascular structures—defined here as vessels with diameters below a fraction of the imaging wavelength (typically $<$ 0.1–0.3$\lambda$). Grade A (high quality) corresponded to images in which these fine vessels were clearly visible and consistently delineated across most of the field of view, forming continuous microvascular patterns. Grade B (medium quality) was assigned when fine vessels were visible only in certain regions, resulting in heterogeneous small-scale vascular detail. Grade C (low quality) was used when fine vessels were largely absent or indistinguishable, and the image was dominated by larger vascular structures.

After consensus between graders, 57 acquisitions were classified as A, 24 as B, and 18 as C. Representative examples of these three quality levels are shown in Fig.~\ref{fig:visual_grading}. Although lower-quality acquisitions exhibit reduced sampling or less favorable imaging conditions, they remain valuable for developing and benchmarking methods intended to operate robustly under realistic constraints.

%% file: sections/Usage_Notes.tex

%% file: sections/Code_Availability.tex
\section{Code Availability}
\label{sec:code_availability}
To support the use of ULMShare, we provide a set of lightweight MATLAB functions for loading raw data together with the associated \texttt{sequence.json} files. An example script illustrating the workflow from raw data to a simple super-resolution map is included. This example relies on existing open-source toolboxes and is intended solely as a demonstration of how to interact with the dataset, not as a reproduction of the ULM maps distributed in the repository.

Full reconstruction of the example ULM maps is not provided, as the internal processing pipeline is hardware-specific, under active development, and not currently suitable for full open-source release. Instead, users are encouraged to rely on established community tools. The \href{https://www.biomecardio.com/MUST/index.html}{MATLAB UltraSound Toolbox (MUST)}~\cite{garcia_simus_2022,cigier_simus_2022, garcia_make_2021} is used for beamforming, and the \href{https://github.com/provostultrasoundlab/TrackingAndLocalizationULM}{Tracking and Localization toolbox}~\cite{leconte_tracking_2024} is used for MB localization and tracking.

In the provided example, clutter is removed using SVD applied to beamformed data to reduce memory usage and computation time. All processing parameters in the example scripts match those summarized in Table~\ref{tab:parameters_ULM}.